\documentclass[prd,showpacs,nofootinbib,preprint,10 pt]{revtex4-1}
\usepackage{amsmath,graphicx,color,xcolor,epsfig}
\usepackage{epstopdf}
\usepackage{hyperref}

\begin{document}
\begin{titlepage}

\begin{center}
{\bf\Large\boldmath{Hadronic Weak Decay $\mathcal{B}_{b}(\frac{1}{2}^+) \to \mathcal{B}(\frac{1}{2}^{+},\; \frac{3}{2}^{+}) +V$}}\\[15mm]
\setlength {\baselineskip}{0.2in}
{\large  Fayyazuddin}\\[5mm]

{\it National Centre for Physics, Quaid-i-Azam University Campus, Islamabad, Pakistan.}\\[5mm]

{\large  M. Jamil Aslam}\\[5mm]

{\it Physics Department, Quaid-i-Azam University, Islamabad, Pakistan.}\\[5mm]
\end{center}

{\bf Abstract}\\[5mm] 
\setlength{\baselineskip}{0.2in} 
It is shown that for the effective Lagrangian with factorization ansatz considered here, the two body hadronic decay  $\mathcal{B}_{b}(\frac{1}{2}^+) \to \mathcal{B}(\frac{1}{2}^{+},\; \frac{3}{2}^{+}) + V$, for $\mathcal{B}_{b}(\frac{1}{2}^{+})$ belonging to the representation $\bar{3}$, only allowed decay channel is $\mathcal{B}_{b}(\frac{1}{2}^+) \to \mathcal{B}(\frac{1}{2}^{+})+ V$, where $\mathcal{B}(\frac{1}{2}^{+})$ belongs to the representation $8$ of $SU(3)$. However, for $\mathcal{B}_{b}(\frac{1}{2}^{+})$ belonging to the sextet representation $6$, the allowed decay channels are $\mathcal{B}_{b}(\frac{1}{2}^+) \to \mathcal{B}(\frac{1}{2}^{+},\; \frac{3}{2}^{+}) + V$, where $\mathcal{B}(\frac{1}{2}^{+})$ and $\mathcal{B}(\frac{3}{2}^{+})$ belongs to the octet representation $8^{\prime}$ and the decuplet $10$ of $SU(3)$, respectively. The decay channel  $\mathcal{B}_{b}(\frac{1}{2}^+) \to \mathcal{B}(\frac{1}{2}^{+}) + V$ is analyzed in detail. The decay rate ($\Gamma$) and the asymmetry parameters $\alpha\;, \alpha^{\prime}\;, \beta\;, \gamma$ and $\gamma^{\prime}$ are expressed in terms of four amplitudes. In particular for the decay $\Lambda_b \to \Lambda + J/\psi$ it is shown that within the factorization framework, using heavy quark spin symmetry, the decay rate and the asymmetry parameters can be expressed in terms of two form factors $F_1$ and $F_{2}/F_{1}$, which are to be evaluated in some model. By using the values of these form factors calculated in a quark model, the branching ratio and the asymmetry parameters $\alpha$ and $\alpha^{\prime}$ are calculated numerically. For other heavy quarks belonging to the triplet and sextet representation, the results can be easily obtained by using $SU(3)$ symmetry and phase space factor. Finally, the decay $\Omega_{b}^{-} \to \Omega^{-} + J/\psi$ is analyzed within the factorization framework. It is shown that the asymmetry parameter $\alpha$ in this particular decay is zero. The branching ratio obtained in the first approximation is compared with the experimental value.
\end{titlepage}

\section{INTRODUCTION}
Heavy flavor physics is of topical interest. New data for decays of $b-$hadrons will be coming out from the LHCb. In 2013 the LHCb Collaboration has performed an angular analysis of the decay $\Lambda_b \to  \Lambda + J/\Psi$ where $\Lambda_{b}$'s are produced in proton-proton $(pp)$ collisions at the centre of mass energy $\sqrt{s} = 7$ TeV at the LHC \cite{LHCb-Collaboration}. By fitting several asymmetry parameters in the cascade decay distribution of $\Lambda_{b}\to \Lambda (\to p \pi^{-}) + J/\Psi(\to \ell^{+}\ell^{-})$, the collaboration has reported the relative magnitude of helicity amplitudes in $\Lambda_{b}\to \Lambda+J/\Psi$ decay and also the transverse polarization of $\Lambda_b$ relative to the production plane. 

Theoretically, the nonleptonic
decay $\Lambda_b \to \Lambda +J/\Psi$ is quite attractive because only factorizable tree diagram contributes to the 
decay and there is no contribution due to $W-$exchange diagrams \cite{2}. In the $b-$baryon sector, the decay $\Lambda_b \to \Lambda +J/\Psi$ has been studied theoretically in quark model by using the factorization hypothesis \cite{3, 4, 5, 6, 7, 8, 9, 10, 11} and the  results of some of these
calculations have been compared to the new experimental results by the LHCb Collaboration. The results of the branching fraction of $\Lambda_b \to \Lambda +J/\Psi$ decay given in the PDG $\mathcal{B}r(\Lambda_b \to \Lambda +J/\Psi)\times \mathcal{B}r(b\to\Lambda_{b}^{0})= (5.8\pm0.8) \times 10^{-5}$ \cite{17} is deduced from the measurements by the CDF \cite{13} and the D$0$ collaborations \cite{14}. The result of branching fraction from the LHCb is still missing for this decay. In the present study, we give a general formalism for $\mathcal{B}_{b}(\frac{1}{2}^+) \to \mathcal{B}(\frac{1}{2}^{+}) +V$, especially with $V=J/\psi$. Using this formalism, we analyze $\Lambda_b \to \Lambda +J/\Psi$ decay in detail.

Heavy baryons with $J^{P}=\frac{1}{2}^{+}$, belong to either representation $\bar{3}$ or the sextet $6$, whereas $J^{P}=\frac{3}{2}^{+}$ belongs only to the the sextet representation of the $SU(3)$\cite{15}:
\begin{eqnarray}
\bar{3}: \ \ \ \ \ \ \ \ \ \ A_{ij}&=&\frac{1}{\sqrt{2}}(q_i q_j - q_j q_i)Q\chi_{M A}\;, \notag \\
6: \ \ \ \ \ \ \ \ \ \ S_{ij}&=&\frac{1}{\sqrt{2}}(q_i q_j + q_j q_i)Q\chi_{M S}\;, \notag \\
6: \ \ \ \ \ \ \ \ \ \ S^{*}_{ij}&=&\frac{1}{\sqrt{2}}(q_i q_j + q_j q_i)Q\chi_{S} \;, \label{1}
\end{eqnarray}
where $q_{i}\; , q_j$ are $u\; , d\; , s$; $Q = b$ or $c$ and $\chi$'s are the spin wave functions \cite{15}. In Eq. (\ref{1}) $A_{ij}\;, S_{ij}$ and $S^{*}_{ij}$ correspond to $J^{P} = \frac{1}{2}^{+}$ and $J^{P}=\frac{3}{2}^{+}$, respectively. The triplet of heavy baryons are
\begin{equation}
(A_{12}\;, A_{13}\;, A_{23}) : (\Lambda_{b\;,\; c}^{0\;,\; +}\;, \Xi_{b\;,\; c}^{0\;,\; +}\;, \Xi_{b\;,\;c}^{-\;,\; 0} )\;,  \label{2}
\end{equation}
whereas the sextet are
\begin{eqnarray}
(S_{11}\;, S_{12}\;, S_{22}) &:& (\sqrt{2}\Sigma_{b\;,\;c}^{+\;, ++}\;,\;\Sigma_{b\;,\; c}^{0\;, \;+}\;, \sqrt{2}\Sigma_{b\;,\;c}^{-\;,\; 0} )\;,  \notag \\
(S_{13}\;, S_{23}) &:& (\Xi^{\prime 0\;,\; +}_{b\;,\;c}\;, \Xi^{\prime -\;,\; 0}_{b\;,\;c})\;,  \notag \\
S_{33}&:& \sqrt{2}\Omega_{b\;,\; c}^{-\;,\; 0}\label{3}.
\end{eqnarray}
In the Standard Model (SM) two body hadronic decays of heavy flavor mesons and baryons are analyzed in terms of the effective Lagrangian or Hamiltonian. Here, we  take the Hamiltonian
\begin{equation}
H_{eff} = V_{cb}V^{*}_{cs}[a_1(\bar{s}c)_{V-A}(\bar{c}b)_{V-A}+a_2(\bar{c}c)_{V-A}(\bar{s}b)_{V-A}]\;, \label{4}
\end{equation}
where $a_1 = C_{1}+\zeta C_{2}$ and $a_2 = C_{2}+\zeta C_{1}$ , with $\zeta$ being the parameter for the possible number of colors. In terms of the diagrams, $a_{1}$ and $a_{2}$ correspond to the contribution from tree and color suppressed tree diagrams, respectively.

In the factorization ansatz, for the tree diagram and color suppressed tree diagram, the relevant matrix elements are $\langle \mathcal{B}_c|(\bar{c}b)_{V-A}|\mathcal{B}_b\rangle$ and $\langle \mathcal{B}_s|(\bar{s}b)_{V-A}|\mathcal{B}_b\rangle$, respectively. First, one can notice that that $\bar{c}b$ is $SU(3)$ singlet, whereas $\bar{s}b$ is $SU(3)$ triplet. Now
\begin{eqnarray}
3 \times \bar{3} &=& 8 +1\;, \notag \\
3 \times 6 &=& 10 +8^{\prime}\;. \label{5}
\end{eqnarray}
Hence the possible decay modes for $\mathcal{B}_b(\frac{1}{2}^{+})$, for the first term in Eq. (\ref{4}) are
\begin{eqnarray}
\bar{3}: \; \;\;\   \ (\Lambda_{b}^{0}\;, \Xi_{b}^{0}\;, \Xi_{b}^{-} ) &\to & (\Lambda_{c}^{+}\; , \Xi_{c}^{+}\; , \Xi_{c}^{0})(D_s^{-})^{\ast}\; , \notag \\
6:  \; \;\;\  (\Sigma_{b}^{+}\; ,\; \Sigma_{b}^{0}\; ,\; \Sigma_{b}^{-}) &\to & (\Sigma_{c}^{++}\; ,\; \Sigma^{+}_{c}\; ,\; \Sigma_{c}^{0})^{*}(D_{s}^{-})^{*} \; ,\notag \\
(\Xi_{b}^{\prime 0}\; , \;\Xi_{b}^{\prime -}) &\to & (\Xi^{+}_{c}\; ' \;\Xi^{0}_{c})^{*}(D_{s}^{-})^{*} \; ,\notag \\
\Omega_{b}^{-} &\to & \Omega_{c}^{0}(\Omega^{*0}_{c}) (D_{s}^{-})^{*} \; .\label{6}
\end{eqnarray}
Some of these decays given in Eq. (\ref{6}) have been studied in ref. \cite{19}. The main focus of the present study is the heavy to light decays of $b$-baryons.

For the color suppressed tree diagram, as noted in Eq. (\ref{5}), for $\mathcal{B}_{b}(\frac{1}{2}^{+})$ belonging to the representation $\bar{3}$ the possible decay mode is
\begin{equation}
\mathcal{B}_{b}(\frac{1}{2}^{+}) \to \mathcal{B}({\frac{1}{2}}^{+}) J/\Psi \; ,\label{7}
\end{equation}
with $B({\frac{1}{2}}^{+})$ belongs to the octet representation $8$ of $SU(3)$. However, for $\mathcal{B}_{b}(\frac{1}{2}^{+})$ belonging to the sextet representation, we have two possible decay modes:
\begin{eqnarray}
\mathcal{B}_{b}(\frac{1}{2}^{+}) &\to & \mathcal{B}({\frac{1}{2}}^{+}) J/\Psi \; ,\notag \\
\mathcal{B}_{b}(\frac{1}{2}^{+}) &\to & \mathcal{B}({\frac{3}{2}}^{+}) J/\Psi \; .\label{8}
\end{eqnarray}
For this case, $\mathcal{B}(\frac{1}{2}^{+})$ and $\mathcal{B}(\frac{3}{2}^{+})$ belong to the octet representation $8^{\prime}$ and decuplet representation $10$ of $SU(3)$, respectively. 
For the decay $\mathcal{B}_{b}(\frac{1}{2}^{+}) \to \mathcal{B}({\frac{1}{2}}^{+}) J/\Psi $, the decay channels are
\begin{eqnarray}
\bar{3}: \;\;\; (\Lambda_{b},\; \Xi_{b}^{0},\; \Xi_{b}^{-}) &\to & (\Lambda,\; \Xi^{0},\; \Xi^{-}) J/\Psi \; ,\notag \\
6:\;\;\;(\Sigma_{b}^{0},\; \Sigma_{b}^{-},\; \Sigma_{b}^{+}) &\to & (\Sigma^{0},\; \Sigma^{-},\; \Sigma^{+}) J/\Psi \; , \notag \\
(\Xi_{b}^{0\prime},\; \Xi_{b}^{-\prime})  &\to & (\Xi^{0},\; \Xi^{-}) J/\Psi \; ,\label{9}
\end{eqnarray}
where $\Lambda\;, \Xi^{0}\;, \Xi^{-}$ are members of the octet representation $8$ and $\Sigma^{0},\; \Sigma^{-},\;\Sigma^{+},\;\Xi^{0},\;\Xi^{-}$ are members of the octet representation $8^{\prime}$. This study focus on the analysis of  $\mathcal{B}_{b}(\frac{1}{2}^{+}) \to \mathcal{B}({\frac{1}{2}}^{+}) V$ decays.

For the decay $\mathcal{B}_{b}(\frac{1}{2}^{+}) \to \mathcal{B}({\frac{3}{2}}^{+}) J/\Psi$, where $\mathcal{B}_{b}(\frac{1}{2}^{+})$ belong the representation $6$, the decay channels are
\begin{eqnarray}
(\Sigma_{b}^{0},\; \Sigma_{b}^{-},\; \Sigma_{b}^{+}) &\to & (\Sigma^{* 0},\; \Sigma^{* -},\; \Sigma^{* +}) J/\Psi \; , \notag \\
(\Xi_{b}^{0\prime},\; \Xi_{b}^{-\prime})  &\to & (\Xi^{* 0},\; \Xi^{* -}) J/\Psi \; ,\label{9a}\\
\Omega^{-}_{b} &\to & \Omega^{-} J/\Psi \; .\notag 
\end{eqnarray}
where the last decay is most interesting in this category.

\section{Hadronic Weak Decay of Baryon $\mathcal{B}_{b}(\frac{1}{2}^{+}) \to \mathcal{B}({\frac{1}{2}}^{+}) V$: A General Formalism}
For the decay
\begin{equation}
\mathcal{B}_{b}(\frac{1}{2}^{+})(p) \to \mathcal{B}({\frac{1}{2}}^{+})(p^\prime)+ V(k,\epsilon)\label{11}
\end{equation}
where $p = p^{\prime}+k$ and $k \cdot \epsilon = 0$, the Lorentz structure of the $T-$ matrix is given by
\begin{equation}
T=\frac{1}{(2\pi)^{9/2}}\sqrt{\frac{m m^{\prime}}{2 p_0 p^{\prime}_0 k_0}}\bar{u}(p^{\prime})[\gamma \cdot \epsilon (A(s)+B(s)\gamma_{5})+ i \epsilon^{\mu}\sigma_{\mu \nu}k^{\nu}(C(s)+D(s)\gamma_{5})]u(p)\; .\label{12}
\end{equation}
In Eq. (\ref{12}) the amplitudes $A,\;B,\;C$ and $D$ are the function of the square of momentum transfer, i.e.,  $s = (p-p^{\prime})^2$.
In the rest frame of baryon $\mathcal{B}_{b}$
\begin{eqnarray}
m &=& p^{\prime}_0+k_0 \; ,\notag \\
\vec{p}^{\; \prime} &=&- \vec{k} = -|\vec{k}|\vec{n}\; . \label{13}
\end{eqnarray}
In this particular frame, one can write
\begin{equation}
T = \chi^{\dagger}_{f} M \chi_{i} \label{14}
\end{equation}
where
\begin{equation}
M = \frac{1}{(2\pi)^{9/2}}\frac{1}{\sqrt{2k_0}}[i f_1\vec{\sigma}\cdot (\vec{n}\times \vec{\epsilon})+g_1\vec{\sigma}\cdot \vec{\epsilon}+f_2 \vec{n}\cdot \vec{\epsilon}+g_2 (\vec{n}\cdot \vec{\epsilon})(\vec{\sigma}\cdot \vec{n})].\label{15}
\end{equation}
with $\vec{\sigma}$ are the Pauli matrices. The amplitudes $f_{1,\; 2}$, $g_{1,\; 2}$ and $h$ can be written in terms of $A\;, B\;, C\;, D$:
\begin{eqnarray}
f_1 & = & \frac{|\vec{k}|}{\sqrt{2p^{\prime}_0(p^{\prime}_0 + m^{\prime})}}[A(s)-C(s)(m + m^{\prime})] \; ,\label{16a} \\
g_1 & = & -\frac{1}{\sqrt{2p^{\prime}_0(p^{\prime}_0 + m^{\prime})}}[B(s)(p^{\prime}_0 + m^{\prime})+D(s)(k_{0}(m+ m^{\prime})-m^2_{V})]\; ,\label{16b} \\
f_2 & = &  \frac{1}{\sqrt{2p^{\prime}_0(p^{\prime}_0 + m^{\prime})}}\frac{|\vec{k}|}{k_0}[A(s)(m+m^{\prime})-C(s)m_{V}^2]\; ,\label{16c} \\
g_2 & = & \frac{1}{\sqrt{2p^{\prime}_0(p^{\prime}_0 + m^{\prime})}}\frac{|\vec{k}|^2}{k_0}[-B(s)+D(s)(m + m^{\prime})]\; ,\label{16d} \\
h &=& g_1 + g_2 = \frac{-1}{\sqrt{2p^{\prime}_0(p^{\prime}_0 + m^{\prime})}}\frac{1}{k_0}[B(s)((m+m^{\prime})k_0 - m_{V}^2)+D(s)m_{V}^{2}(p^{\prime}_0 + m^{\prime})]\; .\label{18}
\end{eqnarray}

Under space reflection $\vec{\sigma} \to \vec{\sigma}$, $\vec{n} \to -\vec{n}$ and $\vec{\epsilon} \to -\vec{\epsilon}$, thus $f_1$ and $f_2$ are the parity conserving i.e., $p-$wave amplitudes whereas $g_1$ and $g_2$ are the parity violating $s-$wave amplitudes. We also note that for the transverse polarization of $V$ meson, only $f_1$ and $g_1$ are relevant, whereas, for the longitudinal polarization the relevant amplitudes are $f_2$ and $h$. 
The decay width of the above mode is given by
\begin{equation}
d\Gamma = (2\pi)^{7}\delta^4(p-p^{\prime}-k)\big[\frac{1}{2}Tr(MM^{\dagger}\big]d^{3}p^{\prime}d^{3}k\label{19}
\end{equation}
which gives
\begin{equation}
\Gamma = \frac{|\vec{k}|p_{0}^{\prime}}{2\pi m}\big[2(|f_1)|^2+|g_1|^2)+\frac{k_0^{2}}{m_{V}^2}(f_2)|^2+|h|^2)\big]\label{20}
\end{equation}
The first term on the left hand side of Eq. (\ref{20}) corresponds to the transverse polarization and the second term to the longitudinal one.

Let $\vec{S}$ and $\vec{s}$ be the polarizations (spins) of $\mathcal{B}_{b}$ and $\mathcal{B}$, respectively. The decay probability in terms of these polarization vectors is given by
\begin{equation}
dW = (2\pi)^{7} \delta^{4}(p-p^{\prime}-k)\frac{1}{2}Tr\big[(1+\vec{\sigma}\cdot \vec{s})M(1+\vec{\sigma}\cdot \vec{S})M^{\dagger}\big]d^{3}p^{\prime}d^{3}k\; . \label{21}
\end{equation}
Hence, the transition rate is:
\begin{equation}
\frac{dW}{\Gamma} = \frac{d\Omega_{S}d\Omega_{s}}{(4\pi)^2}[1+\alpha \vec{S}\cdot \vec{n} + \alpha^{\prime}\vec{s}\cdot \vec{n} + \beta \vec{s}\cdot(\vec{S} \times \vec{n}) + ((\vec{s}\cdot \vec{n}) (\vec{S}\cdot \vec{n}))(-1+\gamma^{\prime}) + \gamma \vec{s} \cdot (\vec{n}\times (\vec{S} \times \vec{n}))]\; ,\label{22}
\end{equation}
where
\begin{eqnarray}
\alpha & = & 2 Re[-2f_{1}^{*}g_1+(\frac{k_0}{m_V})^{2}f_2^{*}h]p_{0}^{\prime} \frac{|\vec{k|}}{2\pi m \Gamma}\; , \label{23} \\
\alpha^{\prime} & = & 2 Re[2f_{1}^{*}g_1+(\frac{k_0}{m_V})^{2}f_2^{*}h]p_{0}^{\prime} \frac{|\vec{k|}}{2\pi m \Gamma}\; , \label{24} \\
\beta & = & 2 Im[f_{2}^{*}h]p_{0}^{\prime} \frac{|\vec{k|}}{2\pi m \Gamma}\; , \label{25} \\
\gamma & = & (\frac{k_0}{m_V})^{2}[|f_2|^2 - |h|^2]p_{0}^{\prime} \frac{|\vec{k|}}{2\pi m \Gamma}\; , \label{26} \\
\gamma^{\prime} & = & (\frac{k_0}{m_V})^{2}[|f_2|^2 + |h|^2]p_{0}^{\prime} \frac{|\vec{k|}}{2\pi m \Gamma}\; . \label{27} 
\end{eqnarray}

Following comments are in order. For the transverse polarization, the asymmetry parameters are
\begin{equation}
\alpha  =  -\frac{4Re[f^{*}_{1}g_{1}]p^{\prime}_0 k_{0}}{2\pi m \Gamma} = -\alpha^{\prime}\; , \label{28}
\end{equation}
whereas in case of the longitudinal polarization
\begin{equation}
\alpha  =  \big(\frac{k_0}{m_V}\big)^{2}\frac{2Re[f^{*}_{2}h]p^{\prime}_0 k_{0}}{2\pi m \Gamma} = \alpha^{\prime}. \label{29}
\end{equation}
It is clear from Eqs. (\ref{25}, \ref{26}) and Eq. (\ref{27}), that $\beta$, $\gamma$ and $\gamma^{\prime}$ are non-zero only for the longitudinal polarization.
For the longitudinal polarization, we get exactly the same result as that in the non-leptonic decay of $\mathcal{B}$ baryon, when $V$ is replaced by pseudo-scalar meson $P$ \cite{16}.

\section{Factorization: Baryon Form Factors}
In the factorization framework, the effective Hamiltonian for the decay $\mathcal{B}_{b}(\frac{1}{2}^{+})\to \mathcal{B}(\frac{1}{2}^{+})+J/\psi$ is given by
\begin{equation}
H_{eff} = \frac{G_{F}}{\sqrt{2}} V_{cb}V^{*}_{cs}a_2 \langle 0 |\bar{c}\gamma^{\mu} (1-\gamma_5) c | J/\Psi\rangle \langle \mathcal{B} |\bar{s}\gamma_{\mu}(1-\gamma_5)b | \mathcal{B}_b \rangle \; .  \label{30}
\end{equation}
The relevant matrix elements are
\begin{eqnarray}
 \langle 0 |\bar{c}\gamma^{\mu} (1-\gamma_5) c | J/\Psi\rangle & = &(\frac{1}{2\pi})^{3/2}\frac{1}{\sqrt{2k_0}}F_{J/\Psi}m_{J/\Psi} \epsilon^{\mu}\; ,\label{30a} \\
 \langle \mathcal{B} |\bar{s}\gamma_{\mu}(1-\gamma_5)b | \mathcal{B}_b \rangle & = & (\frac{1}{2\pi})^{3}\sqrt{\frac{m m^{\prime}}{p_0 p^{\prime}_0}} \bar{u}(p^{\prime})[(g_V(k^2)-g_A(k^2)\gamma_{5})\gamma_{\mu}\notag \\
 && - i (f_V(k^2) + h_A(k^2)\gamma_5)\sigma_{\mu \nu}k^{\nu}-(h_V(k^2) - f_A(k^2)\gamma_{5})k_{\mu}u(p) \label{31}]
\end{eqnarray}
where $f_V(k^2)\;, g_V(k^2), f_A(k^2)\;, g_A(k^2)\;, h_V(k^2)$ and $h_A(k^2)$ are the form factors. Now using Eqs. (\ref{30a}) and (\ref{31}) in Eq. (\ref{30}), the $T-$matrix can be written as
\begin{equation}
T=\frac{1}{(2\pi)^{9/2}}\sqrt{\frac{m m^{\prime}}{2 k_0 p_0 p^{\prime}_0 k_0}}\frac{G_{F}}{\sqrt{2}} V_{cb}V^{*}_{cs}a_2 F_{J/\Psi}m_{J/\Psi} \bar{u}(p^{\prime})[\gamma \cdot \epsilon (g_V(k^2)+g_A(k^2)\gamma_{5})- i \epsilon^{\mu}\sigma_{\mu \nu}k^{\nu}(f_V(k^2)+h_A(k^2)\gamma_{5})]u(p). \label{T-matrix}
\end{equation}
Hence, comparing Eq. (\ref{12}) and Eq. (\ref{T-matrix}), one gets
\begin{eqnarray}
A & = & G^{\prime}m_{J/\psi} F_{J/\psi} g_V(k^2)\;,\notag \\ 
B & = & G^{\prime}m_{J/\psi} F_{J/\psi} g_A(k^2)\;, \notag \\
C & = & - G^{\prime}m_{J/\psi} F_{J/\psi} f_V(k^2)\;,\notag \\
D & = & - G^{\prime}m_{J/\psi} F_{J/\psi} h_A(k^2)\;, \label{32}
\end{eqnarray}
where
\begin{equation}
G^{\prime} =V_{cb}V^{*}_{cs}\frac{G_{F}}{\sqrt{2}}(C_2+\zeta C_1)\;.
\end{equation}
The short distance QCD effects are taken care of in the Wilson Coefficients $C_1$ and $C_2$. The long distance interactions are shifted to the form factors $g_V$, $g_A$, $f_V$ and $h_A$ which are needed to be evaluated in some model. 
Using the heavy quark spin symmetry, one can relate the different form factors \cite{Mannel:1990vg} for which there are two choices:
\begin{eqnarray}
(i):\ \ \ \ \ \ \ \ \ \ g_V(k^2) = g_A(k^2) = F_{1}(k^2)+\frac{m_b}{m}F_{2}(k^2)\;, \notag \\
\ \ \ \ \ \ \ \ \ \ \ \  f_V(k^2) = h_A(k^2) = \frac{1}{m}F_2(k^2) \notag \\
(ii):\ \ \ \ \ \ \ \ \ \ g_V(k^2) = - g_A(k^2) = F_{1}(k^2)+\frac{m_b}{m}F_{2}(k^2)\;, \notag \\
\ \ \ \ \ \ \ \ \ \ \ \  f_V(k^2) = - h_A(k^2) = \frac{1}{m}F_2(k^2)\;, \label{33}
\end{eqnarray}
where $m = m_{\Lambda_b}$ and $m_b$ is the mass of $b-$quark which in this work is taken to be to be $4.65$GeV. Thus in terms of the form factors $F_{1}(k^2)$ and $F_2(k^2)$, we can write
\begin{eqnarray}
A & = & G^{\prime}m_{J/\psi} F_{J/\psi} F_{1}(k^2)\bigg(1+\frac{m_b}{m}\frac{F_{2}(k^2)}{F_{1}(k^2)}\bigg) = \pm B\;, \notag \\ 
C & = & - G^{\prime}m_{J/\psi} F_{J/\psi}  F_1(k^2)\frac{1}{m}\frac{F_2(k^2)}{F_1(k^2)} = \pm D.\label{34}
\end{eqnarray}
where $\pm$ sign in Eq. (\ref{34}) corresponds to the choices $(i)$ and $(ii)$, respectively. 
We need the form factors at $k^2 = m^2_{J/\psi}$:
\begin{equation}
F_1( m^2_{J/\psi}) \equiv F_1 \; , \frac{F_2( m^2_{J/\psi})}{F_1( m^2_{J/\psi})} \equiv \frac{F_2}{F_1}\;. \label{35}
\end{equation}
From Eqs. (\ref{16a} - \ref{16d}) by using Eq. (\ref{34}) and Eq. (\ref{35}), we can express the amplitudes $f_1$, $g_1$, $f_2$ and $h$ in terms of form factors $F_1$ and $F_2/F_1$ as
\begin{eqnarray}
f_1 & = & R \frac{|\vec{k}|}{\sqrt{2p^{\prime}_0(p^{\prime}_0 + m^{\prime})}}F_1\big[1+(m_b+(m + m^{\prime}))\frac{1}{m}\frac{F_2}{F_1}\big]\;, \label{36a} \\
g_1 & = & R \frac{p^{\prime}_0 + m^{\prime}}{\sqrt{2p^{\prime}_0(p^{\prime}_0 + m^{\prime})}}F_1\big[\mp1+\big(\frac{\mp m_b(p^{\prime}_0 + m^{\prime})\pm k_{0}(m+m^{\prime})-m^2_{J/\Psi}}{(p^{\prime}_0 + m^{\prime})}\big)\frac{1}{m}\frac{F_2}{F_1}\big]\;, \label{36b} \\
f_2 & = &R \frac{|\vec{k}|}{k_0}\frac{m+m^{\prime}}{\sqrt{2p^{\prime}_0(p^{\prime}_0 + m^{\prime})}}F_1\big[1+\frac{m_b(m+m^{\prime})+m^2_{J/\Psi}}{m+m^{\prime}}\frac{1}{m}\frac{F_2}{F_1}\big]\;, \label{36c} \\
h & = &R \frac{(m + m^{\prime})k_{0}-m^2_{J/\Psi}}{\sqrt{2p^{\prime}_0(p^{\prime}_0 + m^{\prime})}}F_1\big[\mp 1+\big(\frac{\mp m_b(k_{0}(m+m^{\prime}-m^2_{J/\Psi})\pm m^2_{J/\Psi}(p^{\prime}_{0}+m^{\prime}))}{(k_{0}(m+m^{\prime})-m^{2}_{J/\Psi})}\big)\frac{1}{m}\frac{F_2}{F_1}\big]\;, \label{36d}
\end{eqnarray}
with $R=G^{\prime}m_{J/\Psi}F_{J/\Psi}$ which is a dimensionless parameter.

We now consider the decay  $\Lambda_b \to \Lambda + J/\psi$ which is of experimental interest. In order to calculate this decay, various models to evaluate the form factors have been considered in the literature \cite{9, 10, 11}. In reference \cite{6} form factors were evaluated in a quark model, and their values are $F_1  \approx -0.219$ and $F_2/F_1 \approx 0.169$. After putting $F_2/F_1 \approx 0.169$ and other input parameters in Eqs. (\ref{36a}, \ref{36b}, \ref{36c}) and  Eq. (\ref{36d}), the numerical values of the amplitudes are given in Table \ref{Table-I}. These results can be extended for other baryons, by using physical masses for relevant parameters and $SU(3)$ symmetry. 

\begin{table*}[tbp]
\caption{\sf {Numerical values of the amplitudes for $\Lambda_b \to \Lambda + J/\psi$ for, $F_2/F_1 \approx 0.169$. Here $R^2 =(V_{cb}V^{*}_{cs}\frac{G_{F}}{\sqrt{2}}(C_2+\zeta C_1)m_{J/\Psi}F_{J/\Psi})^2 \approx 18.97 \times 10^{-14}  (C_2 + \zeta C_1)^2 $. The values of the masses are used from \cite{17}. Here $'-'$ and $'+'$ signs are for the choices $(i)$ and $(ii)$, respectively.}}
\label{Table-I}
\begin{tabular}{|c|c|}
\hline
Amplitudes & Numerical Values \\ 
\hline
$f_1$  & $%
R\;F_1 (0.644)$   \\ 
\hline
$g_1$  & $%
R\;F_1(\mp 0.880) $  \\ 
\hline
 $f_2$ & $R\;F_1 (1.075)$ \\ 
\hline
 $h$ &$R\;F_1 (\mp 1.197)$\\ \hline \hline%
\end{tabular}%
\end{table*}
Making use of the values of amplitudes outlines in Table \ref{Table-I} ,the value of branching ratio for $\Lambda_b \to \Lambda + J/\psi$ decay is obtained to be (c.f. Eq. (\ref{20}))
\begin{equation}
\mathcal{B}_r \approx 1.18 \times 10^{-2} (C_2 + \zeta C_1)^2\;,  \label{37} 
\end{equation}
where $\zeta = \frac{1}{N_c}$, where $N_c$ is the effective number of colors.  As noted in \cite{6}, there are two regime, viz $N_{c}<1/3$ (Eq. (46)) and large $N_c$ limit. Using the values of Wilson coefficients $C_{2} = -0.257$, $C_{1} = 1.009$ \cite{21} and for different values of $\zeta$ that correspond to the large $N_c$ limit, the values are given in Table \ref{large-NC}. One can see that for $\zeta = 0$, our results of branching ratio is compared with the $8.9\times 10^{-4}$ that is obtained in ref. \cite{11}.
\begin{table*}[ht]
\centering 
\caption{\sf The values of Branching ratio for $\Lambda_{b}\to\Lambda+ J/\psi$ for different values of $\zeta$ with large $N_c$ limit.}
\begin{tabular}{|c|cccc|c|c|c|}
\hline
$\Lambda_b \to \Lambda + J/\psi$ &  $\zeta = 0$  & $\zeta = 0.01$  &  $\zeta = 0.05$ &\\
\hline
 \ \  $Br$ \ \  & \ \  $7.8\times 10^{-4}$ \ \  & \ \  $6.1\times 10^{-4}$ \ \  & \ \  $5.0\times 10^{-4}$ &\\
\hline\hline
\end{tabular}
\label{large-NC}
\end{table*}

Similarly, for the value of $\zeta$ that correspond to small $N_c$ limit, the values of branching ratios are given in Table \ref{small-NC}. The experimental value of the branching ratio \cite{17} is
\begin{equation}
\mathcal{B}_{r}(\Lambda_b \to \Lambda J/\psi)\times \mathcal{B}_{r}(b \to \Lambda^{0}_{b}) = (5.8 \pm 0.8)\times 10^{-5}. \nonumber
\end{equation}
Using $\mathcal{B}(b \to \text{baryon}) \approx 9.29 \times 10^{-3}$,
the experimental value of branching ratio for $\Lambda_{b}\to \Lambda J/\psi$ is $\mathcal{B}_{r} = (6.2 \pm 0.8)\times 10^{-4}$  and it is comparable to our value $6.1\times 10^{-4}$ when $\zeta = 0.48$ as well as for $\zeta = 0.01$.
\begin{table*}[ht]
\centering 
\caption{\sf The values of Branching ratio for $\Lambda_{b}\to\Lambda+ J/\psi$ for different values of $\zeta$ that correspond to small $N_c$ limit.}
\begin{tabular}{|c|cccccc|}
\hline
$\Lambda_b \to \Lambda + J/\psi$ &  $\zeta = 1/3$  & $\zeta = 0.40$  &  $\zeta = 0.45$ &  $\zeta = 0.48$ &  $\zeta = 0.50$& \\
\hline
 \ \  $Br$ \ \  & \ \  $0.74\times 10^{-4}$ \ \  & \ \  $2.6\times 10^{-4}$ \ \  & \ \  $4.6\times 10^{-4}$  \ \  & \ \  $6.1\times 10^{-4}$ \ \  & \ \  $7.2\times 10^{-4}$ &\\
\hline\hline
\end{tabular}
\label{small-NC}
\end{table*}

The values of asymmetry parameters for $\Lambda_b \to \Lambda + J/\psi$ decay are obtained from Eqs. (\ref{28}) and (\ref{29}) and these are
\begin{eqnarray}
\alpha \approx \mp 0.19 \; \; \; ,\; \;  \alpha_{T} \approx \pm 0.39 \; \; \; , \; \; \alpha_{L} \approx \mp 0.58 \notag \\
\alpha^{\prime} \approx \mp 0.98 \; \; \; , \; \; \alpha^{\prime}_{T} \approx \mp 0.39 \; \; \; , \; \; \alpha^{\prime}_{L} \approx \mp 0.58 \label{39}
\end{eqnarray}
The experimental value of the asymmetry parameter $\alpha = 0.18 \pm 0.13$. With our choice $(i)$ of the form factors given in Eq. (\ref{33}), the value of asymmetry $\alpha = -0.19$ is comparable to the values obtained in refs. \cite{3, 4, 5, 6, 7, 8, 9}. However, for choice $(ii)$ of the form factors, the value of asymmetry parameter $\alpha = 0.19$ is comparable to the experimental value $\alpha = 0.18 \pm 0.13$. 

We have discussed $\Lambda_b \to \Lambda + J/\Psi$ decay in detail and with this in hand, for heavy baryon belonging to the representation $\bar{3}$ and $6$, the branching ratio can be easily obtained by using $SU(3)$ symmetry, taking into account the phase space for each baryon decay.
For the decays $\mathcal{B}_{b}(\frac{1}{2}^{+}) \to \mathcal{B}({\frac{1}{2}}^{+}) J/\Psi$, $SU(3)$ gives the relation
\begin{equation}
\bar{3}:\ \ \ \ \ \ (\Xi_{b}^{-}\;, \Xi_{b}^{0}\;, \Lambda_b) \to (\Xi^{-}\;, \Xi^{0}\;, \Lambda)J/\Psi: \ \ \ \ (1\;, 1\;, \sqrt{2/3})\;.\label{39a}
\end{equation}
for $\mathcal{B}_{b}(\frac{1}{2}^+)$ belong to representation $\bar{3}$ and $\mathcal{B}(\frac{1}{2}^+)$ belonging to the octet representation. In case of $\mathcal{B}_{b}(\frac{1}{2}^{+})$ belonging to the sextet representation and $\mathcal{B}(\frac{1}{2}^{+})$ belonging to the representation $8^{\prime}$, $SU(3)$ gives
\begin{eqnarray}
(\Sigma_{b}^{+}\;, \Sigma_{b}^{0}\; \Sigma_{b}^{-}) &\to & (\Sigma^{+}\;, \Sigma^{0}\;, \Sigma^{-})J/\Psi : \ \ \ \ \ \sqrt{2}(-1\;, 1\;, 1)\;, \notag \\
(\Xi^{\prime -}_{b}\;, \Xi^{\prime 0}_{b}) & \to & (\Xi^{-}\;, \Xi^{0})J/\Psi: \ \ \ \ (1\;, 1)\;.\label{39b}
\end{eqnarray}
\section{The Decay $\Omega_{b} \to  \Omega^{-} + J/\Psi$}
In the factorization ansatz, corresponding to the effective Hamiltonian given in Eq. (\ref{4}) the matrix element for $\Omega_{b} \to  \Omega^{-} + J/\Psi$ decay is
\begin{equation}
\mathcal{M} = \frac{G_F}{\sqrt{2}}V_{cb}V^{*}_{cs} (C_2 +\zeta C_1)\langle 0|\bar{c}\gamma^{\mu}(1-\gamma^{5})c|J/\Psi\rangle \langle\Omega^{-}|\bar{s}\gamma_{\mu}(1-\gamma^{5})b|\Omega_{b}^{-}\rangle. \label{41}
\end{equation}
We can write
\begin{equation}
 \langle\Omega^{-}|\bar{s}\gamma_{\mu}(1-\gamma^{5})b|\Omega_{b}^{-}\rangle =\frac{1}{(2\pi)^3}\sqrt{\frac{mm^{\prime}}{p_0 p^{\prime}_0}}[(F_1^{V}-
 \gamma^{5}F_1^{A})(\bar{u}_{\mu}(p^{\prime})u(p))+. . . ]\;.\label{42}
\end{equation}
where dots denote the contribution from other form factors  which are suppressed by a factor of $\frac{1}{m_{\Omega_b}}$ compared to $F_1^{V}$ and $F_1^{A}$ and hence will be neglected. 
From Eq. (\ref{41}) and Eq. (\ref{42}) along with Eq. (\ref{33}), we get
\begin{equation}
|\mathcal{M}|^2 = G^{\prime}F^{2}_{J/\Psi}m^{2}_{J/\Psi}\epsilon^{\mu}\epsilon^{\nu}[u_{\nu}(p^{\prime})\bar{u}_{\mu}(p^{\prime})(F^{V}_{1}-\gamma^{5}F^{A}_1)u(p)\bar{u}(p)(F^{*\; V}_{1}+\gamma^{5}F^{*\; A}_1)]\;, \label{43}
\end{equation}
where $G^{\prime} = \frac{G_F}{\sqrt{2}}V_{cb}V^{*}_{cs}(C_2+\zeta C_1)$. Now
\begin{eqnarray}
\sum_{\text{Polarization}}\epsilon^{\mu}(k)\epsilon^{\nu}(k)&=&\big(-\eta^{\mu\; \nu}+\frac{k^{\mu}k^{\nu}}{m^2_{J/\Psi}}\big)\;,\notag \\
\sum_{\text{Spin}}u_{\nu}(p^{\prime})\bar{u}_{\mu}(p^{\prime}) &=& =-\frac{\gamma \cdot p^{\prime} +m^{\prime}}{2m}\big[\eta_{\nu\; \mu}-\gamma_{\nu}\gamma_{\mu}+\frac{i}{3m^{\prime}}(\gamma_{\nu}p^{\prime}_{\mu}-p^{\prime}_{\nu}\gamma_{\mu})-\frac{2}{3m^{\prime\; 2}}p^{\prime}_{\nu}p^{\prime}_{\mu}\big]\;,\notag \\
\bar{\sum_{\text{spin}}}u(p)\bar{u}(p) &=&\frac{1}{2}\frac{\gamma \cdot p+m}{2m}\;.\label{44}
\end{eqnarray}
Using above equations, the decay rate is given by
\begin{equation}
\Gamma = \frac{1}{2\pi m}|\vec{k}|(G^{\prime}F_{J/\Psi}m_{J/\Psi})^2\big(1+\frac{1}{3}\frac{m^2}{m^{\prime\; 2}}\frac{|\vec{k}|^2}{m^2_{J/\Psi}}\big)\big[|F^{V}_{1}|^2(p^{\prime}_0+m^{\prime})+|F^{A}_{1}|^2)(p^{\prime}_{0}-m^{\prime})\big](C_2+\zeta C_1)^{2}\;.\label{45}
\end{equation}
In particular, for $\Omega^{-}_{b} \to \Omega^{-} + J/\Psi$, we have $m = m_{\Omega_{b}}$ and $m^{\prime} = m_{\Omega}$. Now
\begin{eqnarray}
|\Omega^{-} \rangle&=& \frac{1}{\sqrt{3}}(s^{\uparrow}s^{\uparrow}s^{\downarrow}+s^{\uparrow}s^{\downarrow}s^{\uparrow}+s^{\downarrow}s^{\uparrow}s^{\uparrow}\rangle\nonumber\\
|\Omega_{b}^{-} \rangle&=& -\frac{1}{\sqrt{6}}|s^{\uparrow}s^{\downarrow}b^{\uparrow}+s^{\downarrow}s^{\uparrow}b^{\uparrow}-2s^{\uparrow}s^{\uparrow}b^{\downarrow}\rangle\nonumber\
\end{eqnarray}
In NRQM, relevant operators to $\mathcal{O}(v^2/c^2)$ are (for details see \cite{FD-2017}) $\beta$ and $\beta \sigma_{i}$ with $i = z$. Using $\beta|b\rangle = |s\rangle$, we have
\begin{equation*}
\beta |\Omega_{b}^{-},\frac{1}{2}\rangle = -\frac{1}{\sqrt{6}}|(s^{\uparrow}s^{\downarrow}s^{\uparrow}+s^{\downarrow}s^{\uparrow}s^{\uparrow}-2s^{\uparrow}s^{\uparrow}s^{\downarrow})\rangle
\end{equation*}
and
\begin{equation*}
\beta\sigma_{z} |\Omega_{b}^{-},\frac{1}{2}\rangle = -\frac{1}{\sqrt{6}}|(s^{\uparrow}s^{\downarrow}s^{\uparrow}+s^{\downarrow}s^{\uparrow}s^{\uparrow}+2s^{\uparrow}s^{\uparrow}s^{\downarrow})\rangle.
\end{equation*}
This gives $F^{V}_{1}=0$ and $F^{A}_{1}(0) = -\frac{2\sqrt{2}}{3}$ \cite{18}. Thus
\begin{equation}
\Gamma (\Omega^{-}_{b} \to \Omega^{-}+J/\Psi) \approx 1.76\times 10^{-14} |F^A_{1}|^2 (C_2+\zeta C_1)^2 \text{GeV}\;.\label{46}
\end{equation}
Hence the branching ratio
\begin{equation}
Br(\Omega^{-}_{b} \to \Omega^{-} + J/\Psi) = \frac{\tau_{\Omega_b}}{\hbar}\Gamma (\Omega^{-}_{b} \to \Omega^{-} + J/\Psi) = 2.94 \times 10^{-2}|F^A_{1}|^2 (C_2+\zeta C_1)^2\label{47}
\end{equation}
where $F^{A}_{1} = F^{A}_{1}(m^2_{J/\Psi})$ and $\tau_{\Omega_b}$ is the decay time of $\Omega_{b}$. Now using
\begin{equation}
F^{A}_{1} = \frac{1}{m_{J/\Psi}}\frac{m_{b}m_{s}}{m_{b}+m_{s}}F^{A}_{1}(0)\approx \frac{m_{s}}{m_{J/\Psi}}F^{A}_{1}(0) \approx 0.152\;, \label{48}
\end{equation}
The form factor $F^{A}_{1}$ at $m_{J/\psi}$ is expected to be smaller than $F^{A}_{1}(0)$. For this purpose, we have introduced a dimensionless phenomenological factor $(\frac{1}{m_{J/\psi}})(\frac{m_{b}m_{s}}{m_{b}+m_{s}})$, where the second factor is the reduced mass of the constituents of $\Omega^{-}_{b}$. Using $F_{1}^{A}\approx 0.152$, the branching ratio is
\begin{equation*}
\mathcal{B}_r \approx 6.8 \times 10^{-4} (C_2+\zeta C_1)^2.
\end{equation*}
Corresponding to the different values of $\zeta$, the value of branching ratio is given in the Table \ref{Omega-value}.
\begin{table*}[ht]
\centering 
\caption{\sf The values of Branching ratio for $\Omega_{b}\to\Omega+ J/\psi$ for different values of $\zeta$}
\begin{tabular}{|c|ccc|cccc|}
\hline
$\Omega_b \to \Omega + J/\psi$ &  $\zeta = 0$  & $\zeta = 0.01$  &  $\zeta = 0.05$ &  $\zeta = 1/3$ &  $\zeta = 0.40$ &  $\zeta = 0.44$ & \\
\hline
 \ \  $Br$ \ \  & \ \  $4.5\times 10^{-5}$ \ \  & \ \  $4.1\times 10^{-5}$ \ \  & \ \  $2.9\times 10^{-5}$  \ \  & \ \  $0.8\times 10^{-5}$ \ \  & \ \  $1.8\times 10^{-5}$ \ \  & \ \  $3.0\times 10^{-5}$ &\\
\hline\hline
\end{tabular}
\label{Omega-value}
\end{table*}

Experimental $\mathcal{B}r(\Omega^{-}_{b} \to \Omega^{-} + J/\Psi)\times \mathcal{B}r(b \to \Omega_{b}) = (2.9^{+1.1}_{-0.8})\times 10^{-6}$ with $\mathcal{B}(b \to \text{baryon}) \approx 9.29 \times 10^{-3}$ \cite{17} gives
\begin{equation}
\mathcal{B}r(\Omega^{-}_{b} \to \Omega^{-}J/\Psi) =(3.12^{+1.1}_{-0.8})\times 10^{-5}\;. \label{50}
\end{equation}
Finally, in this model, the asymmetry parameter is
\begin{equation}
\alpha(\Omega^{-}_{b} \to \Omega^{-}J/\Psi) = 0\;.\label{51}
\end{equation}

To conclude: using the effective Lagrangian together with factorization ans{\"a}tz  the two body hadronic decay  $\mathcal{B}_{b}(\frac{1}{2}^+) \to \mathcal{B}(\frac{1}{2}^{+},\; \frac{3}{2}^{+}) + V$ is calculated. In case of the $\mathcal{B}_{b}(\frac{1}{2}^{+})$ belonging to the representation $\bar{3}$, the only allowed decay channel is $\mathcal{B}_{b}(\frac{1}{2}^+) \to \mathcal{B}(\frac{1}{2}^{+})+ V$, where $\mathcal{B}(\frac{1}{2}^{+})$ belong to the representation $8$ of $SU(3)$.  However, if $\mathcal{B}_{b}(\frac{1}{2}^{+})$ belongs to the sextet representation $6$, the allowed decay channels are $\mathcal{B}_{b}(\frac{1}{2}^+) \to \mathcal{B}(\frac{1}{2}^{+},\; \frac{3}{2}^{+}) + V$ where $\mathcal{B}(\frac{1}{2}^{+})$ and $\mathcal{B}(\frac{3}{2}^{+})$ belong to the octet representation $8^{\prime}$ and the decuplet $10$ of $SU(3)$, respectively. We have analyazed the decay channel  $\mathcal{B}_{b}(\frac{1}{2}^+) \to \mathcal{B}(\frac{1}{2}^{+}) + V$ in detail, where the decay rate $\Gamma$ and the asymmetry parameters $\alpha\;, \alpha^{\prime}\;, \beta\;, \gamma$ and $\gamma^{\prime}$ are expressed in terms of four amplitudes. These amplitudes are written in terms of the transverse and the longitudinal polarization of $V$. This general formalism is then applied to the decay $\Lambda_b \to \Lambda J/\psi$. It is shown that within the factorization framework, using heavy quark spin symmetry, the decay rate and asymmetry parameters can be expressed in terms of two form factors $F_1$ and $F_{2}/F_{1}$, which being the non-perturbative quantities needed to be evaluated in some model. Here, by taking the values of these form factors calculated in the  quark model \cite{6} the branching ratio and asymmetry parameters $\alpha$ and $\alpha^{\prime}$ are obtained numerically. By taking the color factor $\zeta = 0.01$ or $\zeta = 0.48$ , the branching ratio for the decay $\Lambda_b \to \Lambda + J/\psi$ is matchable to the corresponding experimental value. Having worked out $\Lambda_b \to \Lambda + J/\psi$ decay, this formalism can be easily applied to other heavy quarks belonging to triplet and the sextet representation, by using $SU(3)$ symmetry and  the phase space factor. Finally, the decay $\Omega_{b}^{-} \to \Omega^{-} + J/\psi$ is analyzed within the factorization framework. It is shown that the asymmetry parameter $\alpha$ in this particular decay is zero. The branching ratio obtained in the first approximation is compared with the experimental value.


\end{document}